# NON-LINEAR DYNAMIC OF ROTOR-STATOR SYSTEM WITH NON-LINEAR BEARING CLEARANCE

# DYNAMIQUE NON-LINEAIRE D'UN ENSEMBLE ROTOR-STATOR COMPORTANT UN ROULEMENT NON-LINEAIRE AVEC JEU

Jean-Jacques Sinou* and Fabrice Thouverez

Laboratoire de Tribologie et Dynamique des Systèmes UMR CNRS 5513,
Ecole Centrale de Lyon, 36 avenue Guy de Collongue, 69134 Ecully, France.

**ABSTRACT**
The study deals with a rotor-stator contact inducing vibration in rotating machinery. A numerical rotor-stator system including a nonlinear bearing with Hertz contact and clearance is considered. To determine the non-linear responses of this system, nonlinear dynamic equations can be integrated numerically. But this procedure is both time consuming and costly to perform. The aim of this paper is to apply the Alternate Frequency/Time Method and the "path following continuation" in order to obtain the non-linear responses to this problem. Next, orbits of rotor and stator responses at various speeds are performed.

**RESUME**
Une étude portant sur la dynamique non-linéaire d'un système dans les machines tournantes est présentée. Nous considérons un système rotor-stator comportant un roulement non-linéaire avec jeu et contact de Hertz. Afin de déterminer la réponse non-linéaire de ce système, les équations dynamiques non-linéaires peuvent être intégrées numériquement. Cependant, cette procédure est coûteuse en terme de temps de calcul et de ressources. Le but de ce papier est de proposer l'application d'une méthode de balance harmonique pour déterminer la réponse non-linéaire du système. Ainsi, les orbites du rotor et du stator sont obtenus pour différentes vitesses de rotation.
**Mots-clés:** dynamique des systèmes, dynamique des rotors, analyse non-linéaire, roulement avec jeux, contact.

## 1. INTRODUCTION

The motivation of this study comes from vibration problems induced by rotor-stator contact in turbo-machinery. In fact, various types of non-linear phenomena and effects appear such as rotor-stator contact and clearance bearing [1-2]. During the recent years, the understanding of the dynamic behaviour of systems with non-linear phenomena have been developed in order to predict dangerous or favourable conditions and to exploit the whole capability of structures through system using in non-linear range. In general, time-history response solutions of the full set of non-linear equations can determine the vibration amplitudes but are both time consuming and costly when parametric design studies are needed. Due to the fact that such non-linear systems occur in many disciplines of engineering, considerable work has been devoted to development of methods for the approximation of frequency responses. One of the most popular method is the Alternate Frequency/Time domain (AFT) method [3], based on the balance of harmonic components. In this study, a rotor/stator system with bearing, including Hertz contact and clearance is firstly presented. Secondly, the efficiency of both AFT method and path following continuation is demonstrated in order to obtain the non-linear behaviour of a rotor-stator system with bearing, including Hertz contact and clearance; this method allows to save time in comparison with a classical Runge-Kutta integration, by transforming non-linear differential equations into a set of non-linear algebraic equations in terms of Fourier coefficients.

## 2. ANALYTICAL MODEL

### 2.1. Nonlinear contact

In this model, the Hertz theory is considered in order to evaluate contact between the balls and the races [4]. As illustrated in Figure 1, each ball can be located by its angular position $\theta_k$. Then, the radial non-linear contact force generated on the $k^{th}$ ball can be defined as follows:

$$F_{radial}(\Delta_r) = K_H (\Delta_r - \delta)^{3/2} \text{ if } \Delta_r \geq \delta \text{ (contact)}; \quad F_{radial}(\Delta_r) = 0 \text{ otherwise (no contact)} \tag{1}$$

where $\delta$ and $\Delta_r$ are the radial clearances value and the relative radial distance between the inner and the outer races of the $k^{th}$ bearing. $\Delta_r$ can be expressed by considering horizontal and vertical displacement of the inner and outer races of the $k^{th}$ bearing. One has $\Delta_r = \cos(\theta_k)(x_{outer} - x_{inner}) + \sin(\theta_k)(y_{outer} - y_{inner})$. The effective stiffness $K_H$ is the combined stiffness off a ball to inner race and outer race contacts and is defined by [4]:

$$K_H = 1 / \left( 1 / K_i^{3/2} + 1 / K_o^{3/2} \right) \tag{2}$$

The ball-bearing model under consideration in this study has equi-spaced balls rolling on the surfaces of the inner and outer races. When the outer ring is fixed and the shaft rotates, the angle $\theta_k$ changes with time. Then, each ball is located by its angular position $\theta_k = \omega_c t + 2\pi(k-1)/N$. Then the precessional angular velocity $\omega_c$ of the balls is given by $\omega_c = R\omega_r / 2(R + R_i)$ where $\omega_c$, $\omega_r$, $R$, $R_i$, $N$ are the rotational speed of bearing, the rotational speed of rotor, the outer diameter of inner ring, the diameter of balls, and the number of balls, respectively. Next, the global bearing reaction can be obtained by summing all the individual contact expressions of each $k^{th}$ bearing. The total restoring force components in $x$ and $y$ directions are

$$F_{contact/x} = \sum_{k=1}^{N} F_{radial} \cos(\theta_k) \qquad F_{contact/y} = \sum_{k=1}^{N} F_{radial} \sin(\theta_k) \tag{3}$$

### 2.2. Rotor-bearing-stator model

The rotor-bearing-stator system under study has the outer race of the ball bearing fixed to a rigid support and the inner race fixed rigidly to the shaft. A constant vertical radial force acts on the bearing due to gravity. The excitation is due to an of unbalance force which introduces a rotational frequency. The bearing is composed with 16 balls and is modeled as explained previously, by considering the non-linearity due to the Hertz contact with clearance. The complete rotor-bearing-stator behaviour can be represented with the following equations:

$$\begin{aligned} m_s \ddot{x}_s + c_s \dot{x}_s + k_s x_s &= F_{contact/x} & m_r \ddot{x}_r + c_r \dot{x}_r + k_r x_r &= m_e e \omega^2 \cos(\omega t) - F_{contact/x} \\ m_s \ddot{y}_s + c_s \dot{y}_s + k_s y_s &= F_{contact/y} - m_s g & m_r \ddot{y}_r + c_r \dot{y}_r + k_r y_r &= m_e e \omega^2 \sin(\omega t) - F_{contact/y} - m_r g \end{aligned} \tag{4}$$

This non-linear system can be also written as follows

$$\mathbf{M\ddot{x}} + \mathbf{C\dot{x}} + \mathbf{Kx} = \mathbf{f}^{NL} + \mathbf{f} \tag{5}$$

where $\mathbf{x} = \{x_s\ y_s\ x_r\ y_r\}^T$. $\mathbf{M}$, $\mathbf{C}$ and $\mathbf{K}$ are the mass, the damping and the stiffness matrices. $\mathbf{f}^{NL}$ and $\mathbf{f}$ include non-linear terms, gravity and unbalance, respectively.

## 3. NON-LINEAR METHOD

Both the harmonic balance method and the continuation schemes are well-known numerical tools to study non-linear dynamics problems. However, the AFT method seems rarely used in engineering applications, and more particurlarly in system with clearance and hertz contact. The general idea is to represent each time history response by its frequency content in order to obtain a set of equations including balancing terms with the same frequency components, and to start an iterative approach to obtain roots of these equations [3]. In this study, the AFT method is used to find the response solutions of non-linear rotor-bearing-stator equations.

### 3.1. Alternate frequency/time domain method

The non-linear system (5) can be written in the following way

$$\mathbf{M}\ddot{\mathbf{x}} + \mathbf{C}\dot{\mathbf{x}} + \mathbf{K}\mathbf{x} + \mathbf{f}^{NL}(\mathbf{x}, \omega, \tau) - \mathbf{f}(\mathbf{x}, \omega, \tau) = \mathbf{g}(\mathbf{x}, \omega, \tau) = 0 \qquad (6)$$

where $\mathbf{M}$, $\mathbf{C}$ and $\mathbf{K}$ are the mass, damping and stiffness matrices. $\mathbf{f}^{NL}$ is the vector containing non-linear expressions due to the non-linear contact. Setting $\mathbf{x} = \mathbf{x}_i + \Delta\mathbf{x}$, $\dot{\mathbf{x}} = \dot{\mathbf{x}}_i + \Delta\dot{\mathbf{x}}$ and $\ddot{\mathbf{x}} = \ddot{\mathbf{x}}_i + \Delta\ddot{\mathbf{x}}$, the displacements $\mathbf{x}$ and $\Delta\mathbf{x}$ are represented with truncated Fourier series $m$ harmonics:

$$\mathbf{x} = \mathbf{X}_0 + \sum_{i=1}^{m}\left[\mathbf{X}_{2i-1}\cos(i\omega t) + \mathbf{X}_{2i}\sin(i\omega t)\right], \quad \Delta\mathbf{x} = \Delta\mathbf{X}_0 + \sum_{i=1}^{m}\left[\Delta\mathbf{X}_{2i-1}\cos(i\omega t) + \Delta\mathbf{X}_{2i}\sin(i\omega t)\right] (7,8)$$

in which $\mathbf{X}_0$, $\mathbf{X}_{2i-1}$ and $\mathbf{X}_{2i}$, $\Delta\mathbf{X}_0$, $\Delta\mathbf{X}_{2i-1}$ and $\Delta\mathbf{X}_{2i}$ are the Fourier coefficients of $\mathbf{x}$ and $\Delta\mathbf{x}$, respectively.

The number $m$ of harmonic coefficients is selected in order to only take into account the significant harmonics expected in the solution. $(2m+1) \times 4$ linear algebraic equations are obtained:

$$\mathbf{AX} + \mathbf{F}^{NL} - \mathbf{F} + (\mathbf{A} + \mathbf{J})\Delta\mathbf{X} + \mathbf{Q}\Delta\omega = \mathbf{0} \qquad (9)$$

in which $\mathbf{A}$ and $\mathbf{J}$ are the Jacobian matrices associated with the linear and non-linear parts of (6). They are given by

$$\mathbf{A} = diag(\mathbf{K}\ \mathbf{B}_1\ \cdots\ \mathbf{B}_j\ \cdots\ \mathbf{B}_m) \quad \text{with} \quad \mathbf{B}_j = \begin{bmatrix} -\omega^2\mathbf{M} & j\omega\mathbf{C} \\ -\omega j\mathbf{C} & -(\omega j)^2\mathbf{M} + \mathbf{K} \end{bmatrix}, \quad \text{and}$$

$$\mathbf{J} = (\Gamma \otimes \mathbf{I}) \cdot \begin{bmatrix} \ddots & & \\ & \left[\dfrac{\partial \mathbf{f}^{NL}}{\partial \mathbf{x}}\right] & \\ & & \ddots \end{bmatrix} \cdot (\Gamma^{-1} \otimes \mathbf{I}).$$

$\mathbf{F}$ and $\mathbf{Q}$ represent the Fourier coefficients of $\mathbf{f}$, and the Fourier coefficients of the derivative of $\mathbf{g}$ with respect to $\omega$, respectively. $\mathbf{F}^{NL}$ represents the Fourier coefficients vector of the non-linear function $\mathbf{f}^{NL}$. $\mathbf{X}$ and $\Delta\mathbf{X}$ contain the Fourier coefficients and Fourier increments of $\mathbf{x}$ and $\Delta\mathbf{x}$, respectively. $\mathbf{F}^{NL}$ is difficult to directly determine from the Fourier coefficients for many non-linear elements. However $\mathbf{F}^{NL}$ can be calculated by using an iterative process [3]:
$\mathbf{X} \xrightarrow{DFT^{-1}} \mathbf{x}(t) \longrightarrow \mathbf{f}^{NL}(t) \xrightarrow{DFT} \mathbf{F}^{NL}$ where DFT defines the Discrete Fourier Transform. The DFT from time to frequency domain is given by

$$\Gamma_{ij} = \begin{cases} 1/(2m+1) & \text{for } i=1 \\ 1/(2m+1)\cos\left((j-1)i\pi/(2m+1)\right) & \text{for } i=2,4,\ldots,2m \\ 1/(2m+1)\sin\left((j-1)(i-1)\pi/(2m+1)\right) & \text{for } i=1,3,\ldots,2m+1 \end{cases} \quad \text{for } j=1,2,\ldots,2m+1 \tag{10}$$

The error vector $\mathbf{R}$ and the associated convergence are given by

$$\mathbf{R} = \mathbf{AX} + \mathbf{F}^{NL} - \mathbf{F} \tag{11}$$

$$\delta_1 = \sqrt{\mathbf{R}_0^2 + \sum_{j=1}^{m}\left(\mathbf{R}_{2j-1}^2 + \mathbf{R}_{2j}^2\right)} \quad \text{and} \quad \delta_2 = \sqrt{\Delta\mathbf{X}_0^2 + \sum_{j=1}^{m}\left(\Delta\mathbf{X}_{2j-1}^2 + \Delta\mathbf{X}_{2j}^2\right)} \tag{12,13}$$

### 3.2. Path continuation

Usually, the system behavior is of interest over a range of values for at least one parameter (in this study, the considered parameter is the speed of shaft rotation $\omega$). In order to save time and to obtain more easily the solution of the system by considering variations of parameter values, the path following technique [3] can be used. In this study, estimation of the neighboring point is obtained by using the Lagrangian polynomial extrapolation method with four points. So, four points on the solution branch are obtained a priori in order to begin the extrapolation scheme. Any point on the solution branch is represented at $(\mathbf{X}_i, \omega_i)$, $\mathbf{X}_i$ and $\omega_i$ being the Fourier coefficients and the frequency parameter, respectively. The arc length between two consecutive points $(\mathbf{X}_{i+1}, \omega_{i+1})$ and $(\mathbf{X}_i, \omega_i)$ is given by

$$\delta s_{i+1} = \sqrt{(\mathbf{X}_{i+1} - \mathbf{X}_i)^T (\mathbf{X}_{i+1} - \mathbf{X}_i) + (\omega_{i+1} - \omega_i)^2} \quad \text{for } i = 0,1 \text{ and } 2 \tag{14}$$

hen, the arc length parameters are given by

$$S_0 = 0; \quad S_1 = \delta s_1; \quad S_2 = S_1 + \delta s_2; \quad S_3 = S_2 + \delta s_3; \quad S_4 = S_3 + \Delta s \tag{15}$$

and by using the Lagrangian extrapolation scheme, the following estimated point at the distance $\Delta s$ can be defined by

$$\begin{bmatrix}\mathbf{X}_4 & \omega_4\end{bmatrix}^T = \sum_{i=1}^{3}\left(\prod_{\substack{j=0 \\ i\neq j}}^{3}\left(\frac{S_3 - S_j}{S_i - S_j}\right)\right)\begin{bmatrix}\mathbf{X}_i \\ \omega_i\end{bmatrix} \quad \text{for } i = 0,1,\ldots,3 \tag{16}$$

### 4. APPLICATION

The AFT method is applied to the rotor-bearing-stator system defined previously. The value parameters are given in Table 1. Figure 2 illustrates the frequency response of this system obtained by using the AFT method with the path following continuation. The resonance peak is observed near 50.5 Hz. We can see that at frequencies between 11-19 Hz, unbalance and gravity forces are of the same order amplitude, so that the rotor and stator responses are complex, as illustrated in Figure 3 and in Figure 4(a). In order to obtain the non-linear responses for the frequency range 11-19 Hz, computations are performed by using various power harmonics: with 7 or more frequency components, there is no visible difference between the orbits obtained with Runge-Kutta process and AFT method. When reducing the number of harmonics further to six, only the AFT method found a totally different solution. This emphasises the problem of the AFT method: it is therefore a method that in general can only be used if some a priori knowledge about the system is available. The calculation by using the AFT method with 6, 8 and 10 harmonics components needs about 200, 220 and 240 CPU seconds, respectively. The calculation by using the 4[th] –order Runge-Kutta process needs about 1800 CPU seconds.

At frequencies between 30-80 Hz, rotor and stator are always in contact and orbits are circular and the first frequency components are sufficient ($m=1$), as illustrated in Figure 3 and in Figure 4(b). At frequencies between 1-11 Hz, the same behaviour can be observed, and rotor-stator are always in contact due to the gravity effect. So, Figure 5 shows the contact evolution for each ball of the bearing while increasing the rotation speed. At frequencies between 11-19Hz, the rotor-stator contact is a complex phenomenon with a succession of contact and no-contact periods. At frequencies between 50-80Hz, rotor and stator are always in contact. As explained previously, an interesting point is the contact's evolution during the transit phase around 11-19Hz. As illustrated in Figure 5(b-e), complex non-linear behaviours are obtained.

## 5. SUMMARY AND CONCLUSION

The Alternate Frequency /Time domain method and the following path continuation were briefly described. They seem interesting when time history response solutions of the full non-linear equations are both time consuming and costly. Moreover, extensive parametric design studies can be done in order to appreciate the effect of specific parameter variation on the response of non-linear systems. This method was applied to a rotor-bearing-stator system with nonlinear ball bearing including hertzian contact and radial clearance. Complex orbits and evolutions of the local contact between the balls and the raceways were obtained.

|  | Item | Units | Value |
|---|---|---|---|
| $\xi_r$, $\xi_s$ | Damping ratio for the rotor and the stator | - | 0.01 |
| $m_e e$ | Unbalance magnitude | kg.m | 50.e-3 |
| $\delta$ | Clearance | m | 2.e-5 |
| $K_H$ | Radial bearing stiffness | N/m | 10.e+10 |
| $\omega_0^s$, $\omega_0^r$ | Natural frequency of the stator and the stator | rad/s | 150; 500 |
| $g$ | Gravity | m/s² | 9.8 |

Table 1: Numerical model of physical parameters

Valeurs numériques des paramètres physiques

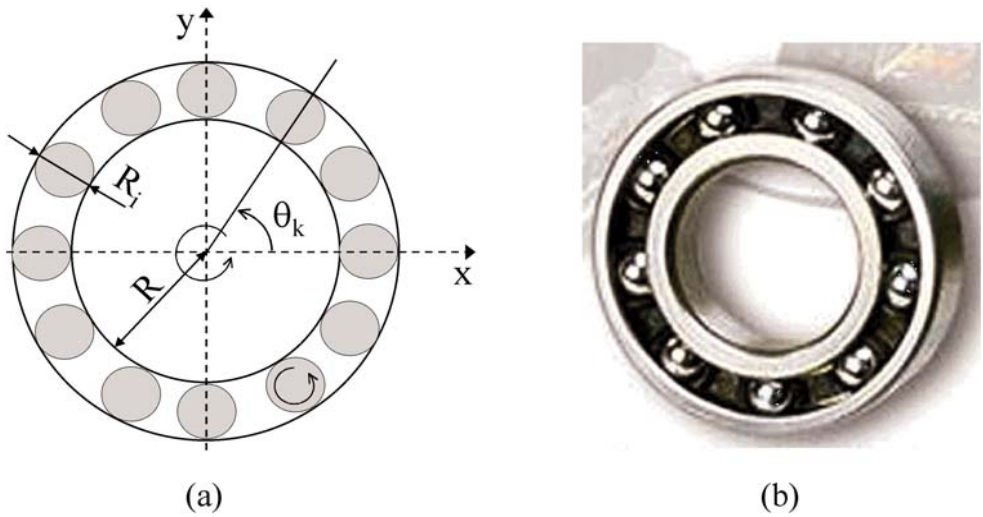

Figure 1: Description of the bearing   (a) location of the $k^{th}$ ball   (b) rolling bearing

Description du roulement   (a) localisation de la $k^{ème}$ bille   (b) roulement à billes

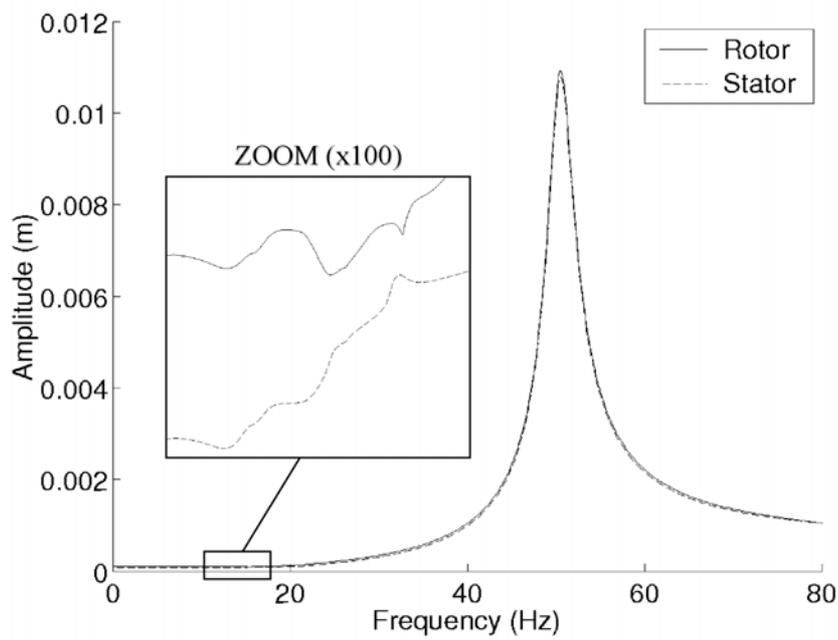

Figure 2: Amplitudes of vibrations versus the rotational frequency

Amplitudes des vibrations par rapport à la vitesse de rotation

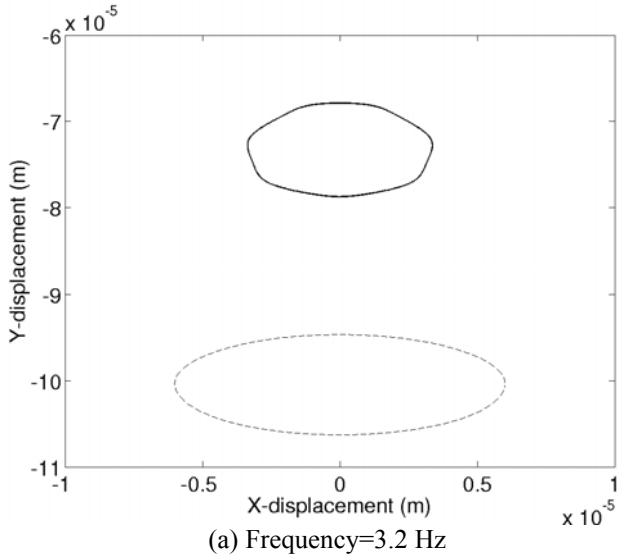
(a) Frequency=3.2 Hz

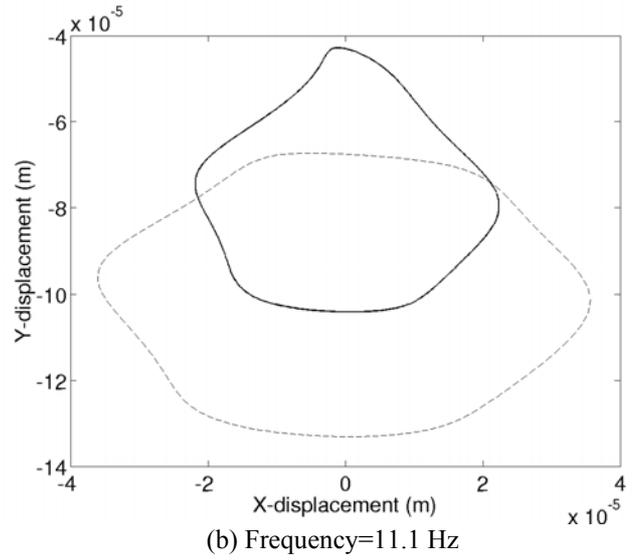
(b) Frequency=11.1 Hz

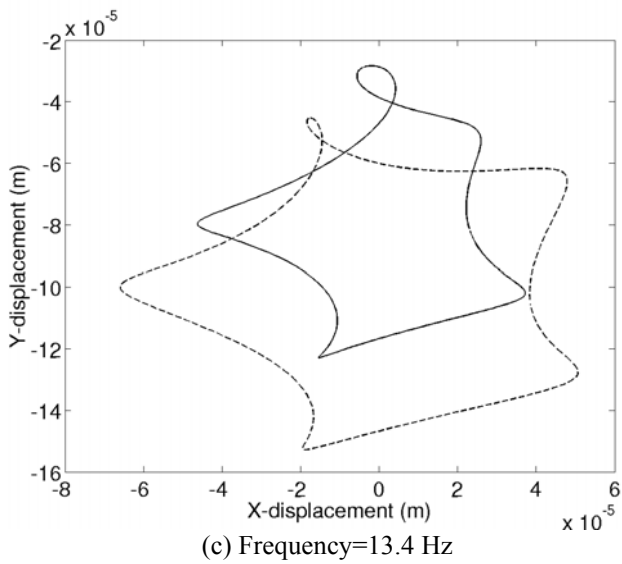
(c) Frequency=13.4 Hz

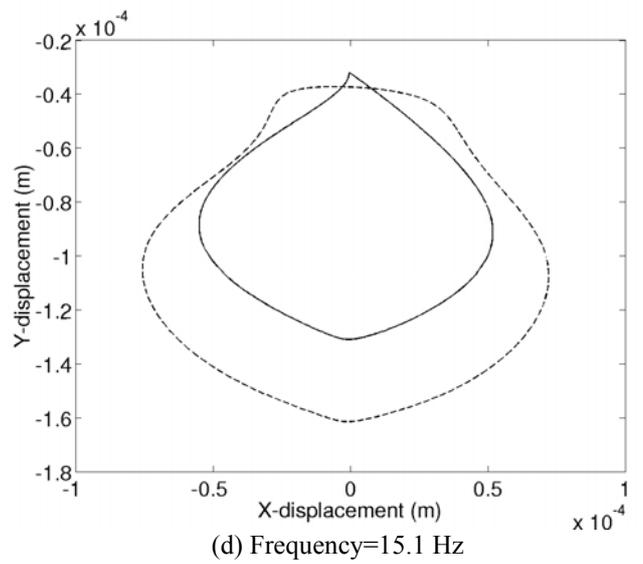
(d) Frequency=15.1 Hz

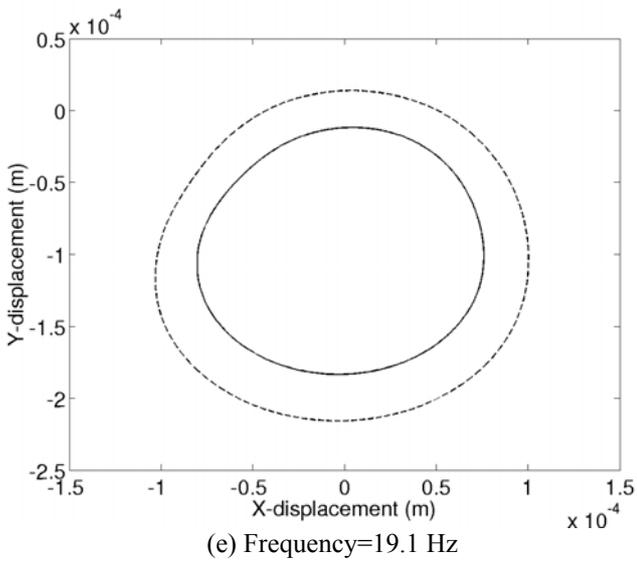
(e) Frequency=19.1 Hz

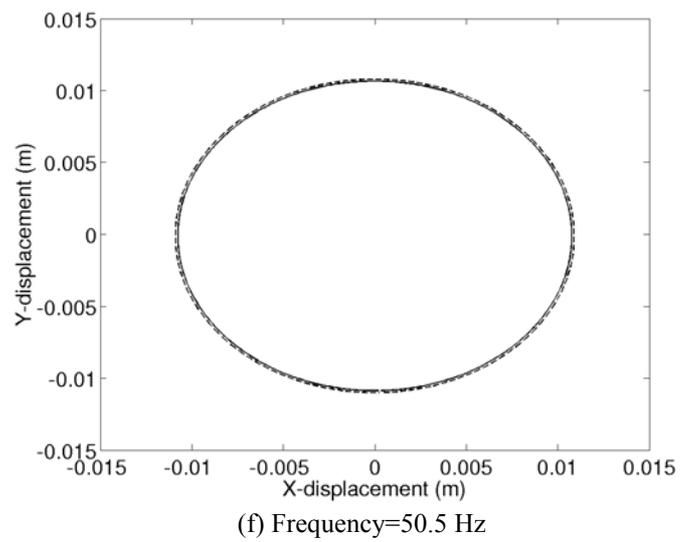
(f) Frequency=50.5 Hz

Figure 3: Orbits of the rotor and the stator at different frequencies
(continuous line: rotor, dashed line: stator)

Orbites du rotor et du stator pour différentes fréquences
(lignes continues= rotor, lignes en pointillés= stator)

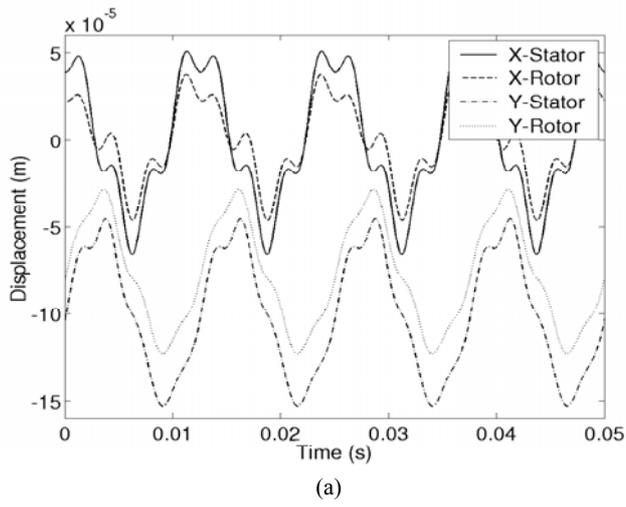 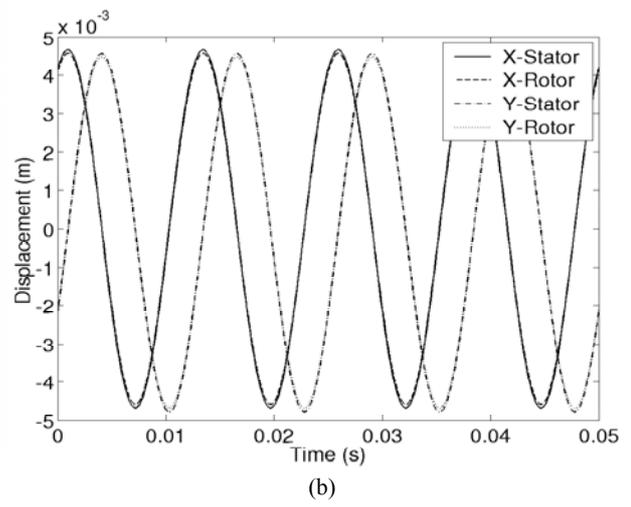

(a)                                           (b)

Figure 4: X,Y – Displacements of the rotor and stator
(a) frequency= 13.4Hz    (b) frequency=47.8Hz

X,Y - Déplacements du rotor et stator
(a) fréquence= 13.4Hz    (b) fréquence=47.8Hz

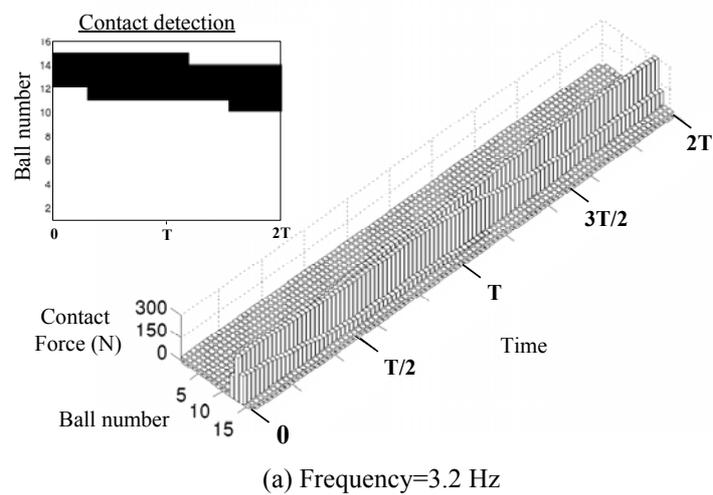
(a) Frequency=3.2 Hz

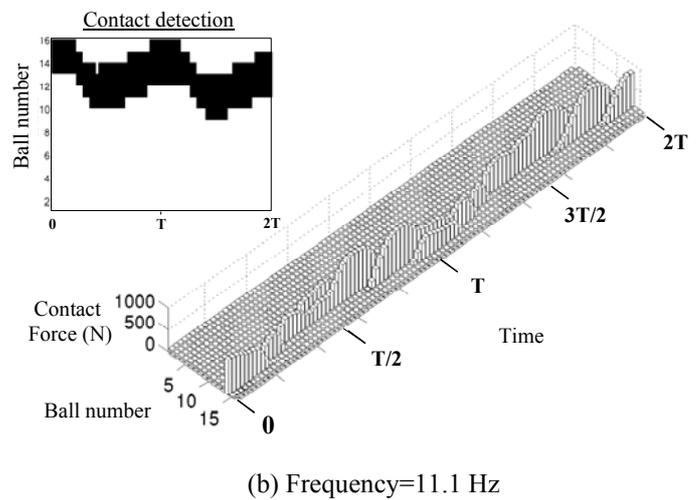
(b) Frequency=11.1 Hz

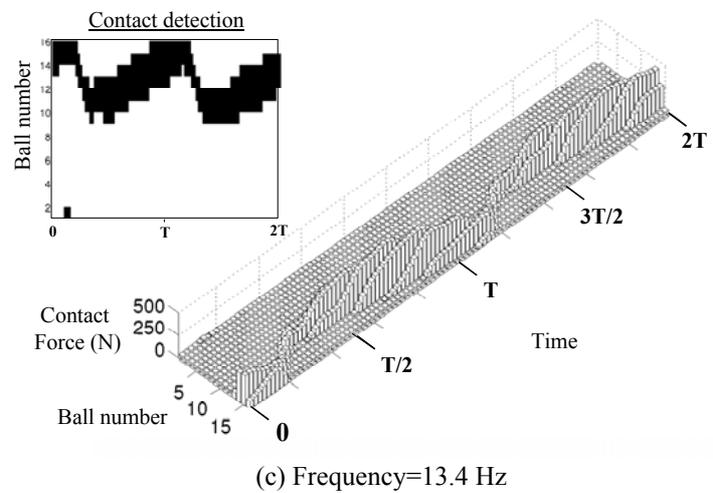
(c) Frequency=13.4 Hz

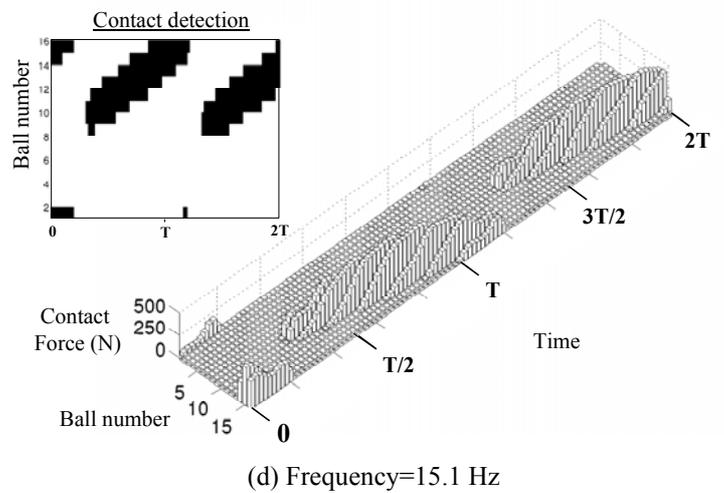
(d) Frequency=15.1 Hz

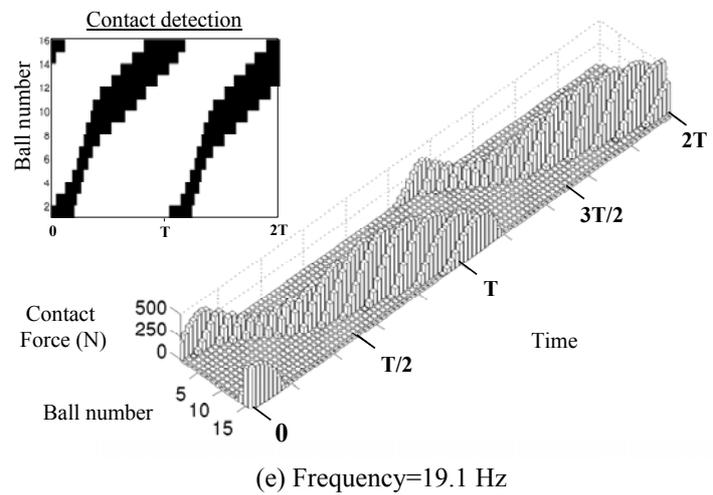
(e) Frequency=19.1 Hz

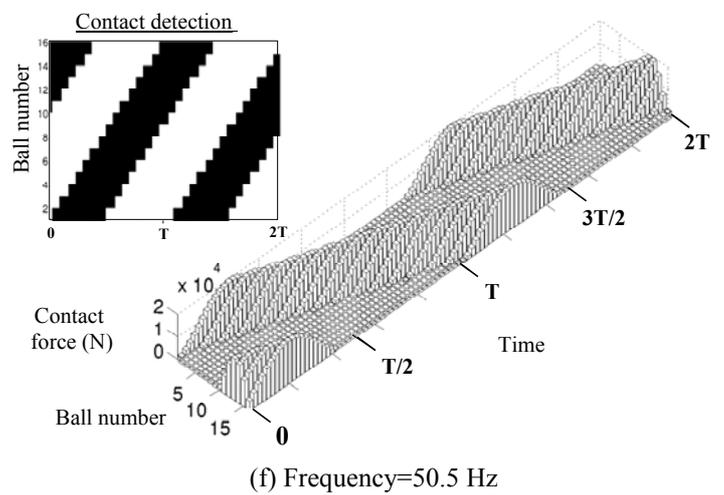
(f) Frequency=50.5 Hz

Figure 5: Evolution of the contact and associated contact force for each ball
(black zone: contact; white zone: non-contact)

Evolution du contact et de la force de contact associée pour chaque bille
(zone noire: contact; zone blanche: pas de contact)